# Lateral spin injection and detection through electrodeposited Fe/GaAs contacts

Sarmita Majumder[1], Bartek Kardasz[1], George Kirczenow[1,2], Anthony SpringThorpe[3] and Karen L. Kavanagh[1]

[1]Dept. Physics, Simon Fraser University, Burnaby, BC V5A 1S6 Canada

[2] Canadian Institute for Advanced Research, Nanoelectronics Program

[3]National Research Council, 1200 Montreal Rd, Ottawa, ON K1A 0R6 Canada

**Abstract**

Efforts to achieve efficient injection of spin-polarized electrons into a semiconductor, a key prerequisite for developing electronics that exploit the electron's spin degree of freedom, have so far met with limited success. Here we report experimental studies of lateral spin injection and detection through electrodeposited Fe/GaAs tunnel contacts. We demonstrate spin injection efficiencies two orders of magnitude higher than for state-of-the-art contacts fabricated via ultra-high-vacuum methods, including those with MgO or $Al_2O_3$ tunnel barriers. To account for this enhancement, we propose that an iron oxide layer that forms at the Fe/GaAs interface during electrodeposition, being magnetic acts as a tunnel barrier with a spin-dependent height, presenting quantum spin transport calculations for such systems. This serendipitous discovery of greatly enhanced efficiency of spin injection into GaAs via electrodeposited contacts introduces a promising new direction for the development of practical semiconductor spintronic devices.

1. **Introduction**

Efficient spin transport into and through a semiconductor is one of the prominent objectives of the field of spintronics [1-6]. Polarized electrons are injected from a ferromagnetic (FM) metal contact (Fe, FeCo, Ni, or Co) [7 – 12] or a magnetic semiconductor hetero-structure [13] and their transport within the semiconductor (Si, Ge, GaAs, or AlGaAs) detected via a non-local spin valve measurement (NLSV). These spin structures typically consist of several FM metal/semiconductor tunnel-junction contacts laterally patterned onto a semiconductor channel, as depicted in figure 1. A bias is applied at one pair of contacts (1 and 3 in figure 1), while two others (4 and 5) develop a potential difference, $\Delta V$, that changes by an amount proportional to the spin polarization in the semiconductor when the in-plane magnetization direction of one of the contacts (3 or 4) is switched. Tunnel junctions between the semiconductor and FM contacts, via a thin heavily-doped $n+$ surface layer, have been found to be crucial for efficient spin injection. [14 – 16] Most spin contacts reported to date have been fabricated within ultra-high-vacuum using molecular beam epitaxy (MBE) [7 – 12] to grow both the semiconductor and the metal film. However, the reported spin voltages, $\Delta V$, have been very small, only 4.0 to 16.8 $\mu$V at low temperatures 10 K to 50 K [9, 10, 17] (applied currents of 1 mA). These signals have been rationalized in terms of phenomenological, classical diffusive transport models [18]. Somewhat higher, but still small, spin voltages (10 $\mu$V, 1 $\mu$A and 12 $\mu$V, 100 $\mu$A ) have been reported when thin MgO [1, 19] or $Al_2O_3$ [2] insulating oxide layers were included between the FM and semiconductor. Others have used electrodeposition to fabricate metal-semiconductor spin devices including Ni/Ge[20] at room temperature and Ni/GaAs[21] at 10 K.

In this work, we have performed lateral spin transport measurements using electrodeposited Fe contacts on similar epitaxial-grown $n+/n$-GaAs (001) substrates. We show that this method, while producing epitaxial Fe/GaAs tunnel junctions, also introduces a much stronger asymmetry in the spin up and spin down tunneling probabilities. This results in a higher spin polarization and therefore much higher spin voltages (4 ± 1 mV at 77 K, for applied currents of 100 $\mu$A) at theelectrodepositedcontacts.

## 2. Experimental Procedures

Our spin test structure (figure 1) consists of 5 parallel bars of Fe/GaAs (001) contacts with dimensions following earlier designs [9]. The 3 inner contact bars are used as injector and detector contacts (each with effective areas 8 $\mu$m × 50 $\mu$m) spaced 4 $\mu$m apart comparable to the GaAs spin diffusion length, $\lambda_D$, (4 $\mu$m at 70 K) [22]. The 2 outer bars are at a much greater distance from the centre (160 μm) to act as normal reference contacts. The electrodeposited contacts were defined via photoresist openings (10 μm × 150 μm) patterned via standard e-beam lithography such that the sides were parallel to <001> crystallographic directions. The substrate consisted of a semi-insulating GaAs (100) bulk substrate with an epitaxial structure consisting of a buffer layer (300 nm undoped GaAs), a Si-doped $n$-GaAs channel (2.5 μm, $3\times10^{16}$ cm$^{-3}$), and a graded-doped layer (15 nm) transitioning to a surface heavily-doped $n^+$ layer (15 nm, $2\times10^{18}$ cm$^{-3}$). Both MBE and MOCVD substrates were grown in well-calibrated systems such that doping concentrations are accurate to within ± 10%. Our own Hall measurements of MOCVD-grown tunnel structure ($2\times10^{18}$ cm$^{-3}$) confirmed a net $n$-type carrier concentration in the channel of $3.9\times10^{16}$ cm$^{-3}$ within the expected range (($2 - 4$) $\times10^{16}$ cm$^{-3}$). Theelectrodepositionwas carried out galvanostatically (15 mA/cm$^2$) at room temperature in an aqueous solution of ferrous sulphate (0.1 M) buffered with ammonium sulphate (0.3 M)(pH 4). [23, 24] The GaAs substrate was the cathode with a Pt rod used as a counter electrode. Prior to deposition, the GaAs was etched in ammonium hydroxide solution (10 % in de-ionized water, 12 sec) and rinsed in de-ionized water. Fe grown via this process nucleates in epitaxial islands where strain relaxation occurs prior to coalescence. Continuous films have bulk-like Fe magnetic properties including a magnetic in-plane inhomogeneity along both <100> and <110> crystallographic directions [25]. To improve the uniformity of the Fe nucleation and growth on the epitaxial layers, ohmic contacts to the doped GaAs layers were made on both ends of the wafer. The GaAs mesa channel was then defined by etching the surface heavily-doped layers in a solution of $H_2O_2$ + $(NH_4)_2SO_4$ +$H_2O$ (1:1:100) to a depth of 60 nm (3 min.). An insulating layer of $SiO_2$ was sputter deposited onto the entire structure and Au pads (200 × 200 µm$^2$) and wire-bonded electrical connections were fabricated using electron-beam evaporation and standard lithography into openings in the $SiO_2$ forming Au/Fe contacts (10 × 50 µm$^2$). Each application of electron-beam resist required baking the sample at 180º C for 60 s. In total, the sample was therefore, annealed for 10 min. at 180º C.

The structural properties were analyzed via high-resolution x-ray diffraction (HXRD) using a monochromated and conditioned Cu K$_{\alpha 1}$ x-ray beam, field-emission (FE) gun transmission electron microscopy (TEM) operating at a voltage of 200 keV, and atom probe tomography via laser ablation of tips fabricated by focussed ion beam milling [26]. Film thickness was measured using field-emission secondary electron microscopy (SEM) of samples cross-sectioned by focused Ga ion beam (FIB) milling. The average growth rate of the electrodeposited Fe film was 480 nm/min [24], with island nucleation and growth that coalesced forming continuous films in 10 ± 1 s. We used growth times of between 10 and 15 s giving film thicknesses of 50 to 100 nm. Analysis of out-of-plane (004) reflections and reciprocal space maps confirmed the films to be highly-oriented (001) with a residual tensile strain of 5×10$^{-5}$ (or 0.005 %), negligible compared to the lattice mismatch (1.4 %). [23, 24] The tensile strain is consistent with a bulk oxygen concentration (< 1 at%) that results in a reduction in the electrodeposited Fe lattice constant by 0.2% and correlates with heterogeneous magnetic properties.[24, 25]

Electrical measurements (current versus voltage) were carried out using a computer-controlled probe station. Magnetization hysteresis curves of blanket films were measured using a SQUID magnetometer with a reciprocating sample option (RSO) where the sample is placed at the center of the SQUID pick-up coil and oscillated by a servo motor during measurements. Magneto-optical Kerr effect (MOKE) was used to image magnetization of patterned Fe bars as a function of applied in-plane field. Spin transport measurements were carried out using a liquid-He cryostat and a computer-controlled electro-magnet capable of producing a variable-field of 0 – 2.3 Tesla, nanovoltmeter and current source (impedance of 10 GΩ).

3. **Results and discussion**

MOKE images of regions from three Fe contact bars labelled 1 to 3 with an applied magnetic field, $H$, sweeping between ± 120 G along a Fe in-plane <100> direction, are shown in figure 2. The area of each image is limited by an aperture within the microscope, diameter 30 μm. The plot is a hysteresis curve showing the degree of polarization, $P$, as a function of $H$. The black arrows indicate the sweep direction of $H$ starting at the top right corner of the $P$ ($H$) curve. The contrast in the image is a measure of change in the in-plane magnetization in the Fe contacts. This change primarily occurs due to in-plane shape anisotropy and other demagnetizing field effects from impurities at the very edges of the contacts. From the hysteresis loop we see that the switching field (red dots on the plot with field value labelled) for the parallel to antiparallel and then anti-parallel to parallel alignment between contacts 1 and 2 occurred at 53 G and -80 G, respectively. The many steps in the hysteresis curves indicate multiple domains are involved in the switching process. There are narrow regions of width approximately 1 $\mu$m at both sides of the bars where the magnetization does not change with the applied field. These correspond to regions of discontinuous film where only discrete islands were present, visible also with plan-view TEM.

Figure 3 (a) shows current density versus voltage (*J-V*) characteristics, as a function of dopant concentration in the semiconductor surface, and type of substrate: bulk (red) or epitaxially-grown GaAs (MOCVD (blue) or MBE (green)). Theelectrodepositedinterfaces follow a rectifying behavior typical of Fermi-level-pinned metal-GaAs contacts, with barrier heights of 0.8 eV based on thermionic emission theory.[23] By increasing the dopant concentration the semiconductor depletion region narrows and tunneling increases dominating the transport, as is true for vacuum-deposited contacts. Higher doping concentrations give larger reverse current, as expected.

Figure 3 (b) compares contact resistance times area, $RA_o$ (log scale), at zero bias as a function of substrate surface doping concentration. Large areaelectrodeposited diodes (triangles), and smaller back-to-back spin contacts on epitaxial substrates (circles), measured at room temperature are compared to data from the literature for MBE-grown Fe/GaAs spin contacts (diamond) measured at 10 K.[27] For $(2.0 \pm 0.5) \times 10^{18}/cm^3$ and $(5.0 \pm 0.5) \times 10^{18}/cm^3$ doping concentrations, $RA_o$ is 422 $\Omega cm^2$ and 10 $\Omega cm^2$, respectively, for large area diodes and 101 and 0.9 $\Omega cm^2$, respectively, for smaller back-to-back diodes, both indicating a 100 times higher contact resistance for the lower doped samples. The inset in figure 3 (b) showing d*I*/d*V* with applied voltage (10 mV to -150 mV) was calculated from figure 3 (a) for the $2 \times 10^{18}$ doped sample. As expected, this is showing a parabolic dependence of the conductance and the minimum is shifted from zero indicating a tunnel behavior [28].

The current density versus voltage characteristics of ourelectrodepositedFe/GaAs contacts (figure 2) indicated that the transport is dominated by tunneling. Furthermore, our contact resistances for similar doping concentration ($5 \times 10^{18}$/cm$^3$) are at least 10 times larger than those reported for MBE-grown Fe/GaAs contacts (0.9 versus 0.1 $\Omega$cm$^2$) even when ours are measured at room temperature while theirs at 10 K [27] indicating the presence of an insulating interfacial layer. The surface of our GaAs substrate is prepared for electrodeposition by a native-oxide etch in ammonium hydroxide. This process is known to leave a residual Ga suboxide with excess As whose thickness decreases with increasing substrate doping concentration.[29, 30] This residual oxide must react with the Fe during electrodeposition while epitaxial Fe is forming. A similar phenomenon has been reported for MBE-grown Fe on intentionally oxidized GaAs [31]. Consistent with this, atom probe tomography detected 4 at.% oxygen within 5 nm from our Fe/GaAs interfaces (data not shown) [26]. If concentrated at the interface this would represent a 0.5 nm thick layer of an Fe oxide phase such as FeO, a sufficient thickness to act as a tunnel barrier. No sulfur was detected by the same atom probe tomography scans although S is present in the electrolyte and is known to adsorb onto GaAs surfaces preventing As interdiffusion.[32]

For NLSV measurements a constant DC current is applied between a central Fe bar (injector) in figure 1 and one of the outer Fe bars while an in-plane magnetic field is swept back and forth. The applied current flows in either a positive or negative direction, such that electrons move either from Fe into the GaAs (positive) at the injector contact or vice versa (negative). Polarized electrons diffuse in all directions from the point of injection at the Fe/GaAs interface including away from the applied current side. The detected spin voltage between a second central bar (the detector) and an outer bar, both ungrounded, is directly proportional to the spin polarization. For parallel magnetic alignment of injector and detector contacts there will be a lower spin resistance path compared to the anti-parallel alignment.

Figures 4 (a) and (b) show the spin voltage versus applied in-plane magnetic field, *H*, at 77 K for applied currents of 100 $\mu$A and -100 $\mu$A, respectively. Forward and reverse directions of field sweeps are indicated by solid (blue) and dotted (red) lines, respectively. The voltage signal is offset by 0.2 mV for the reverse field sweep in figure 4 (a) for clarity. A background voltage of 18 mV has been subtracted from the raw data. The magnetic field is applied along the Fe bars (an in-plane <100> easy axis). For positive current (polarized electrons flow from Fe into the GaAs at the injector contact) a spin voltage of 0.8 ± 0.1 mV is detected in the forward field sweep direction. In the reverse field sweep a smaller peak of 0.1 ± 0.05 mV is seen near -150 G. In figure 4 (b) for negative current (polarized electrons moving from GaAs into Fe at the injector contact), the spin voltage is larger with multiple peaks, from 2 to 4 mV (forward or reverse sweep). Another example of spin voltage from a second set of contacts (background 4.5 mV subtracted) with an applied current of +100 $\mu$A is shown in figure 4 (c). Peaks are detected for both the forward and reverse magnetic field sweeps, with values of 3.0 and 1.0 ± 0.5 mV, respectively.

The asymmetries observed in the spin valve voltages as a function both of the current direction and magnetic field sweep direction are commonly reported by others.[9, 10] Our peak positions are also asymmetric with respect to the nominal zero field. This is due to our electromagnet that is often not exactly at zero field with zero current applied, as its relaxation time varies somewhat from cycle to cycle.

Further verification of spin transport to the detector contact via the channel is obtained from non-local Hanle measurements, where an out-of-plane magnetic field is applied to the same 4 terminal structure. The polarized spins precess around the applied field and with increasing magnitude of the field the spin polarization eventually dephases completely. The spin polarization at the detector as a function of the applied field, $H$, and $x_1$ and $x_2$, the widths of the injector and detector contacts, respectively, can be expressed as [9]:

$$P_{Spin}(H, x_1 x_2) = \left(\frac{inj\ det}{2}\right)\left(\frac{2D\rho}{A}\right) I \int_{w_1}^{w_2} \int_{w_3}^{w_4} \int_0^\infty \frac{e^{-(x_2-x_1)^2/4Dt}}{4\pi Dt} \times e^{-t/\tau_s} \times \cos\left(\frac{g\mu_B Ht}{\hbar}\right) dx_1 dx_2 dt$$

(1)

where, $P_{inj}$, $P_{det}$ are the spin polarizations at the injector and detector, respectively, $I$ the applied current, $A$ the channel cross-sectional area, $\rho$ the resistivity of the GaAs channel, $D$ the spin diffusion coefficient, and $g$ the electron g-factor (-0.44). Diffusive transport (zero drift velocity) and a temperature-independent pre-factor are assumed. $\tau_s$, spin life time is determined from fits to Hanle curves based on this equation.

Non-local Hanle plots are shown in figure 5 (a) obtained at temperatures ranging from 25 K to 77 K. The data shows a peak in each Hanle curve at zero field as expected, and a decay for positive or negative $H$. The solid lines are best fits to each curve using equation (1). The $\tau_s$ found from fits to these curves ranged between 7.8 ± 0.4 ns to 3.2 ± 0.4 ns, 25 K to 77 K, using a chi-squared estimate of the error. $1/(\tau_s)^2$ has been plotted in figure 5 (b) as a function of temperature together with values from the literature for *in situ* MBE [9] Fe/GaAs spin contacts for comparison. Despite the large background signal of our Hanle data, especially for the data at 42 K, the extracted $1/(\tau_s)^2$ follows a linear behavior with similar slope (0.0015 ns$^{-2}$K$^{-1}$) to the MBE data. Both are consistent with a spin-orbit scattering mechanism (Elliot-Yafet) [33] but the intercept is higher ($\tau_s$ is smaller) in our case.

The net spin voltages obtained in our NLSV measurements (0.8 mV to 4 mV, 100 µA and -100 µA at 77 K) are much larger than for Fe/GaAs contacts grown by MBE in UHV (16 µV at 50K) [9] despite an applied current 10 times smaller in our case. Our spin voltages are also two orders of magnitude larger than those that have been reported for MBE grown contacts with MgO or $Al_2O_3$ tunnel barriers between the semiconductor and FM in FM/oxide/Si, Ge, and GaAs junctions. [34-38] The value of our $RA_{spin}$ product, spin resistance ($V_{nonlocal}/I_{applied}$) times channel cross-sectional area, (16 k$\Omega$.µm$^2$) calculated from the spin valve (4 terminal) data is greater than theoretical estimates assuming 100% spin polarization $(RA)_{theo}$ = (4 k$\Omega$.µm$^2$) at 77 K for our channel geometry and concentration. [14, 27] However, the theories in question are phenomenological, make several simplifying assumptions, and also severely underestimate the experimentally observed spin voltages that have been reported by others for the FM/oxide/semiconductor junctions. [34-38] Others have speculated that these discrepancies may be the result of spin accumulation at interfacial states and or to non-uniform tunnel junctions.[35-37] But these hypotheses have yet to be confirmed. Therefore, we focus here on possible reasons why our measured spin voltages are much larger than those observed in experiments on MBE grown systems, including those with oxide barriers; determining why the classical phenomenologies are unable to account satisfactorily for the previous experimental data [34-38] as well as our own, is beyond the scope of the present paper.

We propose that the larger spin voltages and contact resistances that we observe for ourelectrodepositedcontacts are due to an iron oxide interfacial layer that is magnetic and therefore has tunnel barrier heights that are different for majority and minority spin electrons. The thickness of this layer increases for lower semiconductor doping concentrations.[29, 30] For this reason, according to our proposal,electrodepositedcontacts on lightly-doped GaAs tunnel junctions are expected to be more efficient spin filters than are electrodeposited contacts on heavily-doped GaAs. Consistent with this, we do in fact observe larger spin voltages experimentally with electrodeposited contacts on more lightly-doped GaAs (4 mV for $2\times10^{18}$) compared to those on the more heavily-doped GaAs (10 µV for $5\times10^{18}$). Furthermore, as will be shown below by our quantum transport calculations, spin filtering in contacts with spin-dependent tunnel barrier heights can be much more efficient (as evidenced by much stronger spin voltage signals) than in contacts where the barrier height does not depend on the spin, examples of the latter being abrupt Schottky barriers [9, 10] or the non-magnetic MgO [1] or $Al_2O_3$ [2] barriers.

In order to develop a *qualitative* understanding of how measured spin voltages in multi-terminal systems with magnetic tunnel barriers (i.e., spin-dependent barrier heights) at the semiconductor-metal interfaces may differ from those in systems with conventional non-magnetic tunnel barriers we carried out quantum spin transport calculations on small 4-terminal model structures with dimensions of several hundred Angstroms. The electronic and spintronic structures were modeled by the generic tight-binding Hamiltonian:

$$H = \sum_{i\sigma} \varepsilon_{i\sigma} a_{i\sigma}^+ a_{i\sigma} - \sum_{<i,j>\sigma} t_{ij}(a_{i\sigma}^+ a_{j\sigma} + h.c.). \qquad (2)$$

Here $a_{i\sigma}^+$ is the creation operator for an electron with spin $\sigma$ at site *i*. Spin-dependent tunnel barriers as well as electron scattering due to disorder are included in the model through the spin-dependent site energies, $\varepsilon_{i\sigma}$. Nearest neighbor electron hopping is described by the second summation on the right. A total of ~25000 lattice sites were included in the transport calculations. The model system is depicted schematically in the inset of figure 6. It includes two FM electrodes 2 and 3 (dark blue) separated from the semiconductor (orange) by magnetic or non-magnetic tunnel barriers (red). In the FM contacts the majority spin carriers have twice the density of states of the minority spin carriers at the Fermi level. [27] For magnetic tunnel barriers the barrier height for minority carriers was chosen arbitrarily to be 3 times that for majority carriers whereas for non-magnetic barriers the majority and minority spin barrier heights were assumed to be equal. Two non-magnetic electrodes 1 and 4 (pale blue) with non-magnetic tunnel barriers (pink) were also included in the model. Phase breaking of the electronic quantum states and spin relaxation were included in the model by attaching 450 single-channel Büttiker leads [39] at random sites throughout the semiconductor. The inclusion of site disorder through the site energies $\varepsilon_{i\sigma}$, and of the Büttiker leads allows us to model either ballistic or diffusive transport; here we focus primarily on the latter. Quantum transport calculations (see Appendix) were carried out to determine the non-local 4-terminal resistances $R_{\uparrow\uparrow}$ or $R_{\uparrow\downarrow}$ of the model system for parallel or antiparallel magnetizations of the two ferromagnetic contacts, respectively. Here $R_{ij} = V/I$ where the voltage $V$ and current $I$ are indicated in the inset of figure 6. Representative results obtained in this way for magnetic and non-magnetic tunnel barriers, respectively, between ferromagnetic contacts 2 and 3 and the semiconductor are shown in figures 6 (a) and (b) as a function of the electron Fermi energy in the semiconductor. Note that $R_{\uparrow\uparrow}$ and $R_{\uparrow\downarrow}$ are 4-terminal resistances defined as the ratio of the potential difference between contacts 3 and 4 in the inset in figure 6 and the current I flowing through contacts 1 and 2. Because the potential difference is measured between a different pair of contacts than those through which the current flows, $R_{\uparrow\uparrow}$ and $R_{\uparrow\downarrow}$ can be either positive or negative, and both possibilities occur in figure 6. Which occurs for a given set of conditions depends on whether electrons at the Fermi energy are more likely to be transmitted from the semiconductor to contact 3 or to contact 4, and this depends on whether the magnetizations of contacts 2 and 3 are parallel or antiparallel and also on the value of the electron Fermi energy. In figure 6, the calculated spin resistance $|R_{\uparrow\downarrow} - R_{\uparrow\uparrow}|$ that is a measure of the spin voltage signal is larger by factors of approximately 20 to 50 for the case of magnetic tunnel barriers (figure 6 (a)) than for the case of non-magnetic tunnel barriers (figure 6 (b)). In the latter case the difference between $R_{\uparrow\uparrow}$ and $R_{\uparrow\downarrow}$ is solely due to the difference between the minority and majority spin densities of states at the Fermi level in the ferromagnets. While the details of the nonlocal resistance plots shown in figure 6 are sensitive to the microscopic details the model, we find $|R_{\uparrow\downarrow} - R_{\uparrow\uparrow}|$ to be *consistently* much larger for magnetic tunnel barriers than for non-magnetic ones. Significantly, the spin voltage signals observed in our experiments exceed those that have been reported in experiments with non-magnetic MgO or $Al_2O_3$ tunnel barriers [1, 2] by similar factors, i.e., by roughly two orders of magnitude. We conclude that more effective spin filtering by magnetic oxide tunnel barriers than by conventional non-magnetic oxide barriers is capable, in principle, of accounting for the two orders of magnitude enhancement of the spin voltages reported here relative to those observed previously for nonmagnetic oxide tunnel barriers.

We attribute the complex switching behavior of our spin voltage signals as a function of magnetic field that is seen in figure 4, to the successive switching of multiple magnetic domains in each iron electrode, possibly in combination with the switching of multiple magnetic domains in the magnetic iron oxide tunnel barriers. This phenomenon can also be seen in our MOKE data (figure 2). Within the model discussed above the magnetic iron oxide tunnel barrier is also responsible for the very large size of our spin voltage signals relative to those observed in systems with non-magnetic barriers. Modification to spin precession rates and therefore, Hanle peak broadening correlated with smaller spin diffusion times, has also been attributed to local magnetostatic fields originating from the finite roughness of FM film surfaces and at the interface.[35, 36] Our Fe films have a surface roughness of 20 nm which probably contributes to reductions in $\tau_s$. The spontaneous formation of an interfacial magnetic oxide between Fe and GaAs is highly likely at theseelectrodepositedinterfaces and would also have contributed a magnetic noise from domain switching processes.

In summary, we have shown that electrodeposited Fe contacts on GaAs can be used for spin injection and detection. Results from Hanle and NLSV measurements give spin diffusion times (4 to 8 ns) that are a factor 2 lower than those of thinner MBE-grown Fe contacts, but with a 100 times enhancement in the spin voltage signal (4 mV versus 16 µV). We have proposed that these higher spin signals may be related to a thin oxide layer that forms during electrodeposition or contact fabrication and functions as a magnetic tunnel barrier whose height differs for majority and minority spin electrons. We support these experimental results with quantum spin transport calculations that show the strong positive influence of a spin dependent magnetic tunnel barrier on spin transport efficiency at electrodeposited Fe/GaAs contacts. Further experimental and theoretical studies directed at obtaining a more complete understanding of spin transport throughelectrodepositedferromagnet/semiconductor interfaces and through MBE-grown ferromagnet/magnetic oxide/semiconductor heterostructures would be of considerable interest.

**Appendix**

In order to evaluate the quantum spin-transport coefficients for our tight-binding model in the linear response regime, we solved the Lippmann-Schwinger equation for the model Hamiltonian (2) with systems of semi-infinite ideal tight-binding leads connecting the contacts to electron and spin reservoirs and Büttiker leads contacting random semiconductor lattice sites. This yielded the quantum transmission probabilities $T_{\alpha\beta}^{\sigma}$ for electrons at the Fermi energy with spin $\sigma$ between electrodes $\alpha$ and $\beta$; these include the Büttiker leads as well as contacts 1 – 4 shown in the inset of figure 6. This was done for parallel and antiparallel magnetizations of electrodes 2 and 4. We then solved the Büttiker equations [40] for this system numerically to determine the values of the electrochemical potentials applied to contacts 1 – 4 and to the Büttiker leads for which the net electric currents flowing in or out of the voltage contacts 3 and 4 and the Büttiker leads are zero. Finally from the solution of the Büttiker equations we evaluated the non-local 4-terminal resistance of the system $R_{\uparrow\uparrow}$ or $R_{\uparrow\downarrow}$ given by $V/I$ (where the voltage $V$ and current $I$ are indicated in the inset of figure 6) for parallel or antiparallel magnetizations of the two ferromagnetic contacts, respectively. In these calculations the Büttiker leads coupled to the semiconductor between contacts 2 and 3 in such a way as to cause quantum phase breaking but not relaxation of the electron spin polarization, whereas those contacting the semiconductor outside of this region were both phase breaking and spin relaxing. This is consistent with the spin diffusion length in our experimental system being similar to the distance between the two inner ferromagnetic contacts.

**Acknowledgements**

We are grateful to NSERC (grant no. 222869-2009), CMC Microsystems and CIFAR for partial funding of this project, Compute Canada and Westgrid for computational resources, Ty Prosa, (Imago Inc.) for atom probe tomography, Chris Palmstrøm (UCSB) for MBE GaAs epitaxial substrates, and many useful discussions with Palmstrøm and Bret Heinrich (SFU).

**Figure 1**. Schematic diagram of our Fe/GaAs test samples showing 5 Fe contacts (yellow) deposited onto a heavily-doped GaAs surface layer (purple) forming tunnel contacts to a lower-doped GaAs channel (beige).

**Figure 2**. MOKE images and corresponding magnetic polarization, *P*, versus applied in-plane magnetic field, *H*, of regions of three Fe bar contacts, labelled 1, 2, and 3. The size of each image is limited by the objective lens aperture to a diameter of 30 μm. Arrows on the plot indicate the direction of the magnetic field sweeps. Contrast in these images is a direct measure of the change in the in-plane magnetization state. Red dots on the plot show the switching fields for the bars.

**Figure 3** (a). Results from current-density versus voltage characteristics of Fe/GaAs diodes as a function of substrate doping concentration, *n*, bulk (red) and epitaxially grown MBE (green) or MOCVD (blue) for large area diodes (0.8 mm of diameter). (b). Log plot of contact resistance at zero bias times area, versus *n*, large area diodes (0.8 mm of diameter) (triangles); epitaxial back to back spin contacts (8×50 $\mu m^2$)(circles) and literature report for *in situ* MBE Fe/GaAs (diamond). The insert is a plot of dI/dV from the MOCVD epitaxial $2\times10^{18}/cm^3$ data in figure 3 (a).

**Figure 4**. Voltage measured between contacts 4 and 5 in figure 1 versus applied in-plane magnetic field, *H*, for current *I* through contacts 1 and 3 at 77 K for contacts to an MOCVD-grown substrate with a surface dopant concentration of $2\times10^{18}/cm^3$. Figure 4 (a) *I* = 100 μA. Detected voltage is offset by 0.2 mV for the reverse field direction for clarity. Figure 4 (b) *I* = -100 μA. For positive current electrons flow from Fe into the GaAs at the injector contact 3. Forward and reverse direction of *H*-field change is indicated by solid and dotted lines,

respectively. Voltage peaks and dips signal switching of some of the magnetic domains in contacts 3 and 4 to anti-parallel alignment. Results for a second set of contacts measured in the same way with a positive current of 100 µA are shown in figure 4 (c).

**Figure 5** (a). Nonlocal Hanle measurements using four terminals, applying an out-of-plane magnetic field, $H$, as a function of temperature. Solid lines are the calculated fits using the diffusion equation (1). Vertical offsets have been added to the plots for clarity. Figure 5 (b). Plot of $1/(\tau_s)^2$ (extracted from fits to the Hanle data in figure 5 (a) versus temperature for electrodeposited (squares) and MBE (from literature [6]) (triangles) Fe/GaAs (001) spin contacts.

**Figure 6**. Calculated non-local resistances $V/I = R_{\uparrow\uparrow}$ and $R_{\uparrow\downarrow}$ for parallel and antiparallel magnetization of the ferromagnetic contacts 2 and 3 (dark blue), respectively, versus the semiconductor Fermi energy, $E_F$, for the model system shown in the inset. Contacts 1 and 2 are the current leads while 3 and 4 are the voltage leads. $E_F = 0$ is the bottom of the semiconductor conduction band in the absence of disorder. Disorder is, however, included in the model through the site energies, $\varepsilon_{i\sigma}$, enabling transport through the semiconductor at somewhat lower energies. Contacts 1 and 4 (pale blue) and the associated tunnel barriers (pink) are non-magnetic. The Büttiker leads (purple) carry no net current but break the phase of the electron wave functions throughout the semiconductor. They also induce electron spin relaxation except in the region of the semiconductor between contacts 2 and 3. In figure 6 (a) the tunnel barriers (red) between the ferromagnetic contacts 2 and 3 and the semiconductor are magnetic, while in figure 6 (b) they are non-magnetic (as in the case of $Al_2O_3$ barriers) and, accordingly, the calculated spin resistance $|R_{\uparrow\downarrow} - R_{\uparrow\uparrow}|$ is much larger in figure 6 (a) than in figure 6 (b). The differences

between the results in figure 6 (a) and 6 (b) are due entirely to the different barriers in the two cases. The values of the model parameters that describe the ferromagnetic metal in figure 6 (a) and 6 (b) are the same.

Figure 1

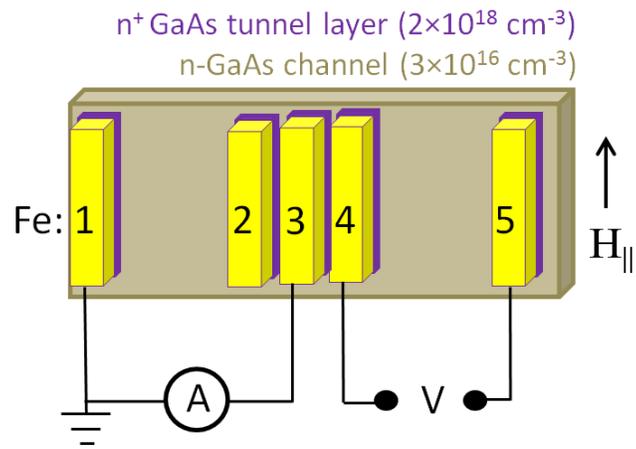

Figure 2

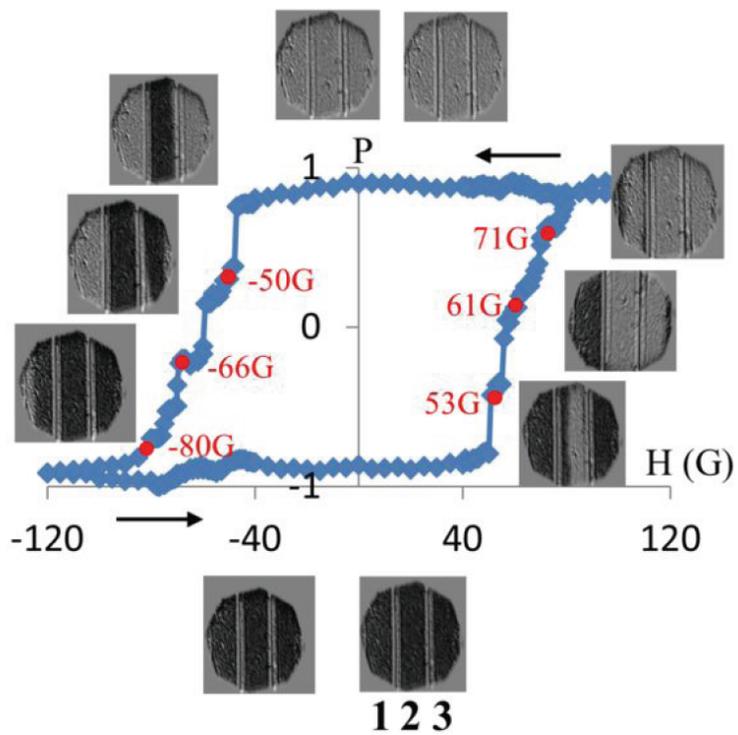

Figure 3

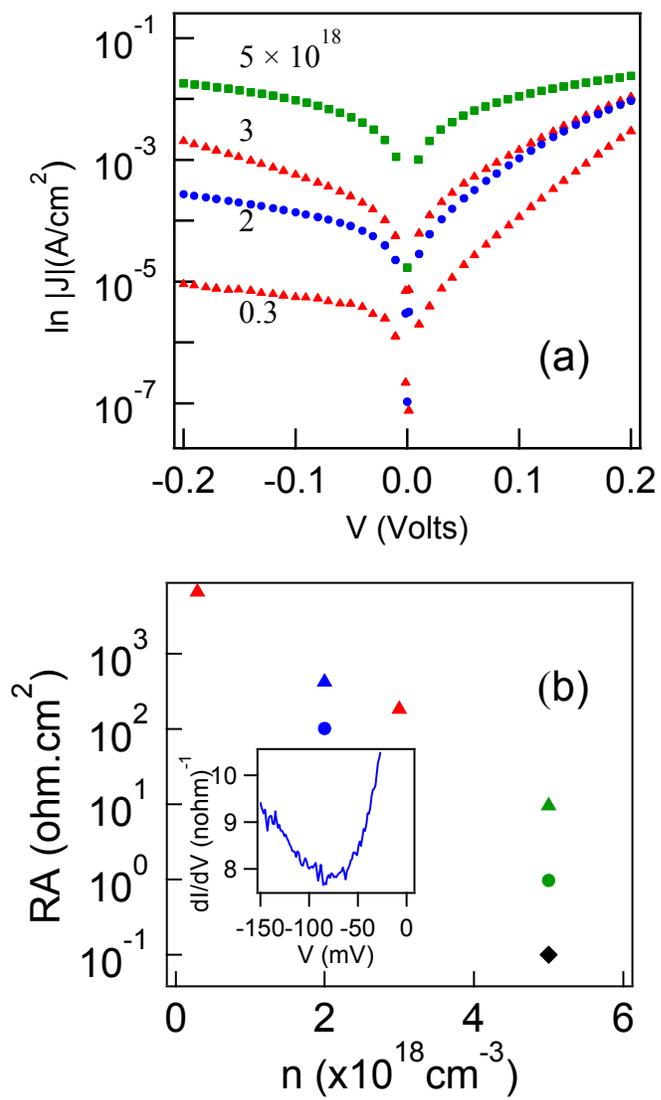

Figure 4

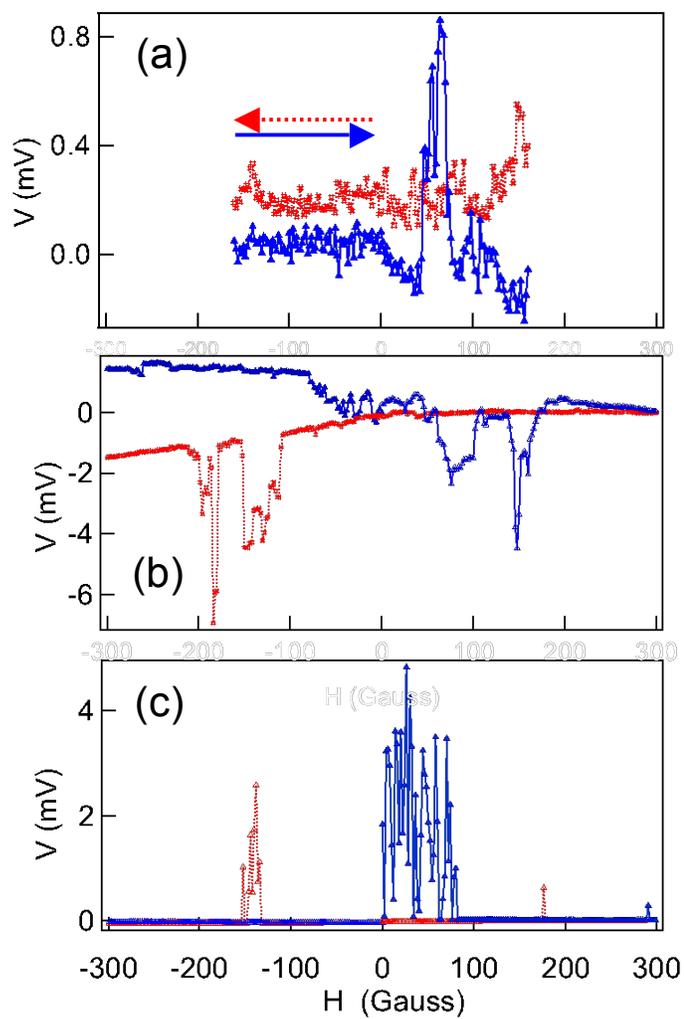

Figure 5

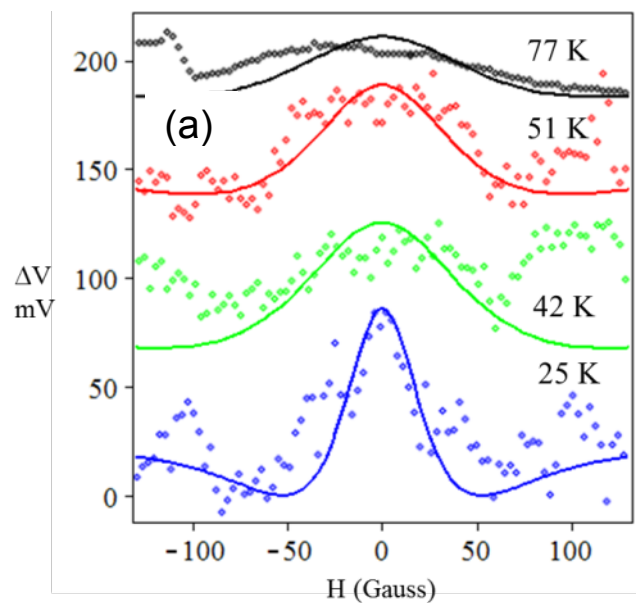

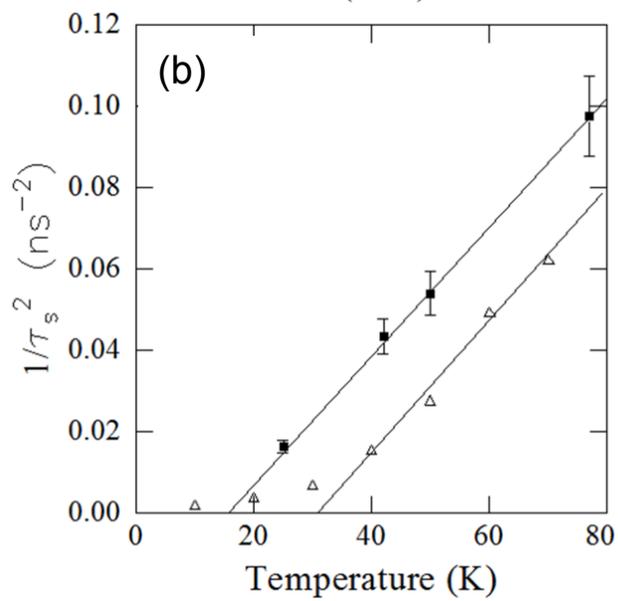

Figure 6

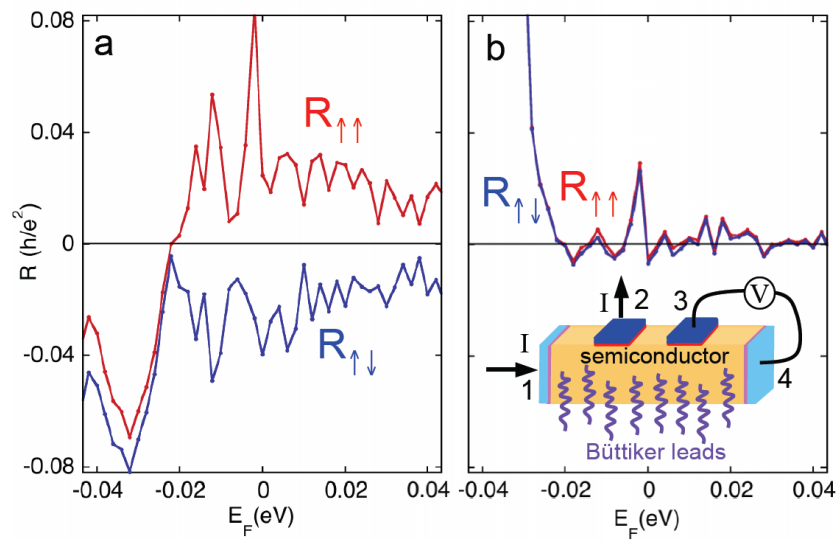